# ADVANCED RADIO RESOURCE MANAGEMENT FOR MULTIANTENNA PACKET RADIO SYSTEMS

Stanislav Nonchev and Mikko Valkama

Department of Communications Engineering, Tampere University of Technology, Tampere, Finland

stanislav.nonchev@tut.fi
mikko.e.valkama@tut.fi

## ABSTRACT

In this paper, we propose fairness-oriented packet scheduling (PS) schemes with power-efficient control mechanism for future packet radio systems. In general, the radio resource management functionality plays an important role in new OFDMA based networks. The control of the network resource division among the users is performed by packet scheduling functionality based on maximizing cell coverage and capacity satisfying, and certain quality of service requirements. Moreover, multi-antenna transmit-receive schemes provide additional flexibility to packet scheduler functionality. In order to mitigate inter-cell and co-channel interference problems in OFDMA cellular networks soft frequency reuse with different power masks patterns is used. Stemming from the earlier enhanced proportional fair scheduler studies for single-input multiple-output (SIMO) and multiple-input multiple-output (MIMO) systems, we extend the development of efficient packet scheduling algorithms by adding transmit power considerations in the overall priority metrics calculations and scheduling decisions. Furthermore, we evaluate the proposed scheduling schemes by simulating practical orthogonal frequency division multiple access (OFDMA) based packet radio system in terms of throughput, coverage and fairness distribution among users. In order to completely reveal the potential of the proposed schemes we investigate the system performance of combined soft frequency reuse schemes with advanced power-aware packet scheduling algorithms for further optimization. As a concrete example, under reduced overall transmit power constraint and unequal power distribution for different sub-bands, we demonstrate that by using the proposed power-aware multi-user scheduling schemes, significant coverage and fairness improvements in the order of 70% and 20%, respectively, can be obtained, at the expense of average throughput loss of only 15%.

## 1. INTRODUCTION

Development of new advanced packet radio systems continues progressively. Relevant work in this direction includes, e.g., 3GPP long term evolution (LTE) [1], WiMAX [2], IMT-Advanced [3] and the related work in various research projects, like WINNER [4]. These developments rely on OFDMA air interface, scalable bandwidth operation, exploitation of MIMO technologies and advanced convergence techniques. The operating bandwidths are divided into large number of orthogonal subcarriers for multi-access purposes and efficient reduction of the effects of inter-symbol interference (ISI) and inter-carrier interference (ICI).The major driving forces behind these developments are the increased radio system performance, in terms of average and peak cell throughputs, low latency and reduced operating expenditures. While the average and peak throughputs are typically emphasized, also the fairness and cell-edge coverage are equally important quality measures of cellular radio systems [5], [6]. Yet another key measure becoming all the time more and more important is the energy consumption of the radio access network.

Another important issue in these deployments is co-channel interference (CCI) that users at the cell borders are experiencing based on low signal to noise ratio (SNR) and high power emitters from neighbouring cells' base stations in their communication channel. Solution for this problem in OFDMA cellular networks is utilization of controlled frequency reuse schemes. Recent literature studies indicate that soft frequency reuse scheme has a capacity gain over the other hard frequency reuse schemes [7]. The overall functionality of frequency reuse schemes is based on applying specific power masks (fraction of the maximum transmission power level) over the whole system bandwidth. In soft frequency reuse (SFR) [8],[9] the frequency band, shared by all base stations (BS) (reuse factor is equal to 1) is divided into sub-bands with predefined power levels as illustrated in Figure 1. Consequently, the users near to and far away from the BS are allocated with different powers, limiting the impact of and to the neighbouring cells.

In general, all user equipments (UEs) within base station (BS) coverage are sharing the available radio resource (radio spectrum). This is controlled by the packet scheduler (PS) utilizing selected scheduling metrics. In both uplink (UL) and downlink (DL), the packet scheduler works in a centralized manner being located at the BS. It is generally fairly well understood that the overall radio system performance, in terms of throughput and fairness, depends heavily on the PS functionality. The PS operation can build on instantaneous radio channel conditions, quality of service (QoS) requirements and traffic situations of the served UEs [5], [6]. In the emerging radio systems, multi-antenna MIMO technologies increase spectral efficiency and also provide the PS with extra degree of freedom by offering a possibility to multiplex data of one or more users on the same physical resource (spatial domain multiplexing, SDM) [10], [11]. Such functionality requires additional feedback information provided by mobile stations (MS) to the base station, typically in terms of channel quality information (CQI) reports. The main disadvantage of MIMO-SDM technique is increased signaling overhead and scheduling complexity. Finally, an additional important aspect in scheduling functionality is the interaction with other radio resource management (RRM) entities, namely fast link adaptation (LA) and reliable re-transmission mechanisms.

Packet scheduling principles are generally rather widely investigated in the literature and many new scheduling algorithms have been proposed, see, e.g., [12]-[16]. Most of them are considering equal BS transmission power distribution among all physical resource blocks (PRBs), which is not necessarily a practical or optimum case as already indicated above. In practice, unequal power allocation between different PRB's can be used, e.g., to control the interference between neighboring cells or sectors in frequency re-use 1 radio systems. Stemming now from our previous work in advanced PS algorithms reported in [15]-[18], we extend the studies here to incorporate energy efficiency and cell power bounds considerations in the scheduling decisions based on different SFR power mask configurations. Our starting point is the advanced modified proportional fair scheduler for MIMO reported in [16], which was shown to improve cell-edge coverage and user fairness compared to state-of-the-art. In this paper, we introduce a new energy- or power-aware scheduling scheme taking into account the applied power pattern within the overall radio spectrum. This is called MIMO power-aware modified proportional fair (MPMPF) scheduler in the continuation. The performance of the proposed scheduler is investigated using both single-user MIMO (spatial multiplexing for individual UE streams) and multi-user MIMO (spatial multiplexing also for streams of different UEs) transmit-receive schemes in 3GPP LTE system context. The used system-level figures of merit are cell throughput distribution, cell-edge coverage and Jain's fairness index [19]. For simplicity and illustration purposes, 1x2 (SIMO) and 2x2 (MIMO) multi-antenna scenarios are assumed in the continuation.

The rest of the paper is organized as follows: Section 2 describes the SFR scheme and power mask configurations, while Section 3 is dedicated to the MIMO scheduling principles and proposed power-aware scheduling scheme. Section 4 gives an overview of the overall system model and simulation assumptions. The simulation results and analysis are presented in Section 5, while the conclusions are drawn in Section 6.

## 2. SOFT FREQUENCY REUSE SCHEMES

In general, SFR scheme reserves part of the frequency band for the cell-edge users and uses the power bound specified for it by the power mask. The rest of the unallocated sub-bands are dedicated to the near to BS users. One practical solution for sub-band division reported in the literature studies is 3 and therefore we are considering it in our case. Generally there is no restriction on the soft reuse factor [7],[8]. The same apply for the choice of the power mask. We have chosen for our evaluation the following power mask configurations: PM1 (0dB, -4dB, -4dB) and PM2 (0dB, -1dB, -4dB). The values in the brackets represent nominal transmission power values in dB as shown in Figure 1. Notice that if the sub-band allocated to the cell-edge UEs is not fully occupied, it can be still used by the other UEs.

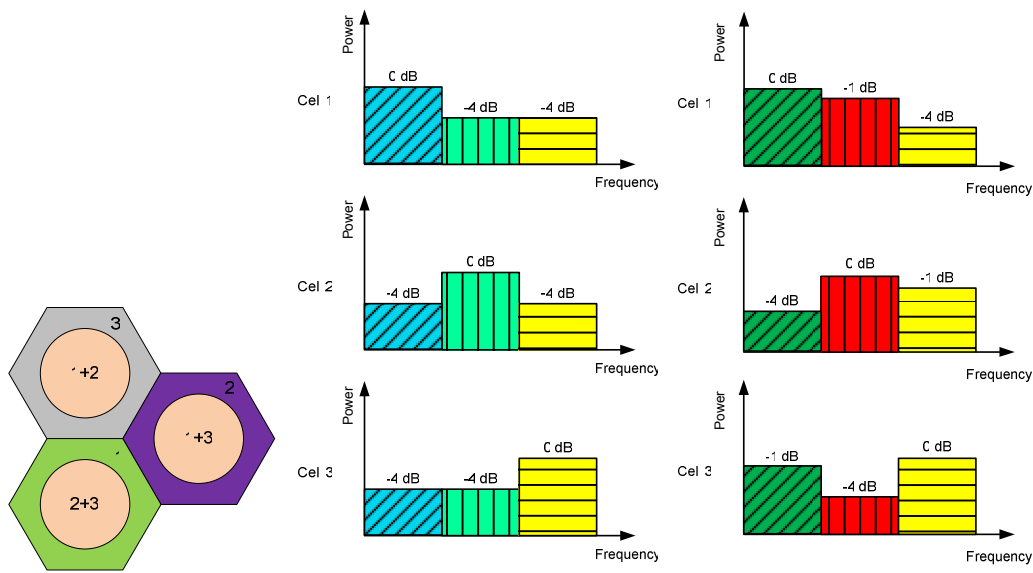

Figure 1: Soft Frequency reuse scheme with 3 sub-bands division.

## 3. POWER-AWARE MIMO SCHEDULING

In general, packet scheduler is part of the overall radio resource management mechanism at BS. Therefore, PS functionality depends heavily on the collaboration with other RRM units such as *link adaptation* (LA) and *Hybrid ARQ* (HARQ) manager, as depicted in Figure 2. The scheduling decisions are based on the selected scheduling metric, utilizing typically the CQI reports and acknowledgements from UEs, per given transmission time interval (TTI) in time domain and per physical resource block (PRB) in frequency domain [5]. Moreover, we assume here that the LA unit can feed the PS with power allocation information on a PRB-per-PRB basis. MIMO functionality, in turn, requires both single-stream and dual-stream CQI feedback by each UE. Consequently, the BS decides whether the particular time-frequency resource is used for transmitting (*i*) only one stream to a specific UE, (*ii*) two streams to a specific UE (SU-MIMO) or (*iii*) 1+1 streams to two different UEs (MU-MIMO) [6]. BS also handles proper transmit power allocation in all the cases (*i*)-(*iii*), respectively, such that target packet error rate is reached with selected modulation and coding scheme (MCS).

### 3.1. Proposed Scheduler at Principal Level

The proposed scheduler developments are stemming from the widely used two-stage (see e.g. [15], [16]) multi-stream PF approach. In the first stage, within each TTI, UEs are ranked based on the full bandwidth channel state information and throughput calculations. This is called time-domain (TD) scheduling step. For MIMO case, different spatial multiplexing possibilities (one-stream, dual-stream SU, dual-stream MU) are taken into account, in calculating all the possible reference throughputs. In the second stage, the scheduling functionality is expanded in frequency-domain (FD) and spatial-domain (SD) where the actual PRB allocation takes place. First the needed PRB's for pending re-transmissions (on one stream-basis only) are reserved and the rest available PRB's are allocated to the selected UE's from the first stage. The actual priority metric in FD/SD stage is evaluated at PRB-level taking into account the available stream-wise channel state information, the transmit power allocations and the corresponding throughput calculations. The exact scheduling metrics are described below.

### 3.2. Scheduling Metrics

Here we describe the actual scheduling metrics used in ranking users in the TD scheduling stage as well as mapping the users to FD/SD resources in the second stage. First a power-aware extension of ordinary multi-stream PF scheduler is described in sub-section 3.2.1, used as a reference in the performance simulations. Then the actual proposed modified power-aware metric is described in sub-section 3.2.2.

#### 3.2.1 Power-Aware Multistream Proportional Fair

Stemming from ordinary multi-stream PF principle, the following power-aware PF (PPF) scheduling metric evaluated at each TTI is deployed

$$\gamma_{i,k,s} = \arg\max_{i} \left\{ \frac{P_{k,s}}{P_{\max}} \frac{R_{i,k,s}(n)}{T_i(n)} \right\} \tag{1}$$

Here $R_{i,k,s}(n)$ is the estimated instantaneous throughput of user *i* at sub-band *k* on stream *s* at TTI *n*. $T_i(n)$ is the corresponding average delivered throughput to the UE *i* during the recent past [16],[17]. $P_{k,s}$, in turn, is the transmission power for sub-band *k* on stream *s* and $P_{max}$ is the maximum transmission power for any sub-band. Notice that the power ratio $P_{k,s}/P_{max}$ has a direct impact on the ranking of the users, since the deployed sub-band power levels affect the corresponding estimated and delivered throughput quantities. The scheduler in (1) is only used as a reference in evaluating the performance of the actual proposed scheduler to be described below.

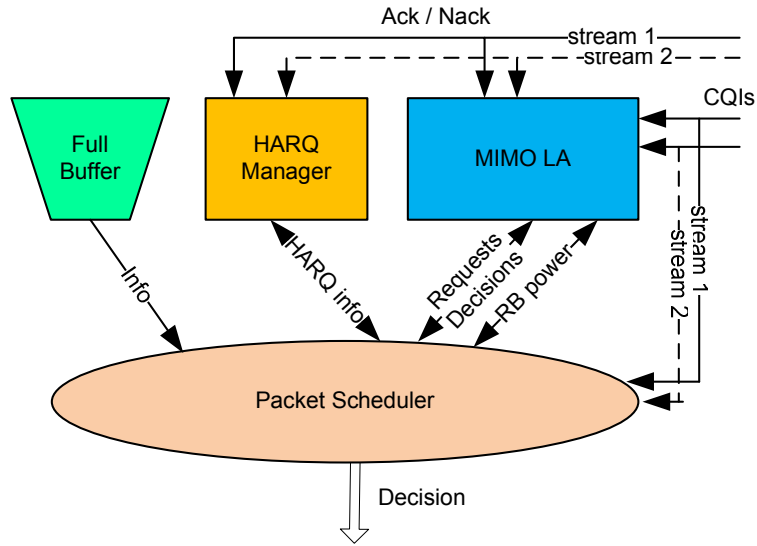

Figure 2: RRM functionalities and scheduling process.

### 3.2.2 Modified Power-Aware Multi-stream Proportional Fair (MPMPF)

Starting from the earlier work in [16], the following modified power-aware multi-stream PF metric is proposed in this paper:

$$\bar{\gamma}_{i,k,s} = \arg\max_{i} \left\{ \frac{P_{k,s}}{P_{\max}} \left( \frac{CQI_{i,k,s}(n)}{CQI_i^{avg}(n)} \right)^{\alpha_1} \left( \frac{T_i(n)}{T_{tot}(n)} \right)^{-\alpha_2} \right\}$$

(2)

In the above metric, $\alpha_1$ and $\alpha_2$ are scheduler optimization parameters ranging basically from 0 to infinity. $CQI_{i,k,s}$ is the CQI of user $i$ at PRB $k$ and stream $s$, and $CQI_i^{avg}$ is the average CQI of user $i$. $T_{tot}(n)$ is the average delivered throughput (during the recent past) to all users ranked in TD stage served by the BS.

The proposed scheduling metric in (2) is essentially composed of three elements affecting the overall scheduling decisions. The first ratio takes into account the transmit power fluctuations in BS for each PRB due to applied SFR scheme. The second ratio is the relative instantaneous quality of the individual user's radio channels over their own average channel qualities in a stream-wise manner. The third ratio is related to measuring the achievable throughput of individual UE's against the corresponding average throughput of scheduled users. The power coefficients $\alpha_1$ and $\alpha_2$ are additional adjustable parameters that can be tuned and controlled to obtain a desired balance between throughput and fairness. This will be illustrated in Section 5.

The basic idea of incorporating the sub-band power ratio $P_{k,s} / P_{max}$ into the scheduling metric (2) is that the sub-band power levels obviously affect the link adaptation and thereon the estimated supportable throughput as well as the actual delivered throughput. Thus by taking the power fluctuations and the transmit power levels in BS for each PRB due to SFR schemes into account, we seek higher fairness in the scheduling between truly realized UE throughputs, and thereon better cell-edge coverage. This will be demonstrated in Section 5. Also since the power is understood in a stream-wise manner, this gives somewhat higher priority to single-stream transmissions which also helps in increasing the coverage.

## 4. RADIO SYSTEM SIMULATIONS

### 4.1. Basic Features

Extensive quasi-static system simulator for LTE downlink providing traffic modeling, multiuser packet scheduling and link adaptation including HARQ is used for evaluating the system level performance of the proposed packet scheduling scheme, following 3GPP evaluation criteria [1]. The 10 MHz system bandwidth case is assumed, being composed of 1024 sub-carriers (out of which 600 are active) and divided into 50 physical resource blocks (PRB) each consisting of 12 sub-carriers with sub-carrier spacing of 15 kHz. Pilot signals are sent from base station to mobile station to determine the instantaneous channel condition. The mobile stations measure the actual channel states and the information is reported to the BS. The actual reported CQI's are based on received signal-to-interference-and-noise ratios (SINR), calculated by the UE's for each PRB. Here the UE's are assumed to use linear MMSE (LMMSE) receivers for MIMO 2x2 system simulation and MRC for SIMO case. Additionally, the UEs always report single-stream SINR as well as both single user (SU) and multi-user (MU) dual-stream SINR's at the corresponding detector output.

In a single simulation run, mobile stations are randomly distributed over standard hexagonal cellular layout with altogether 19 cells each having 3 sectors. As a concrete example, the number of active users in the cell is set to 15 and the UE velocities equal 3km/h. The path losses for individual links are directly determined based on the individual distances between the mobile and the serving base station. On the other hand, the actual fading characteristics of the radio channels are collected each TTI (1ms) and depend on the assumed mobility and power delay profile. Due to the centralized approach statistics are collected only from the central cell site while the others simply act as sources of inter-cell interference. The used MIMO scheme for performance evaluation purposes is per-antenna rate control (PARC) with two transmit antennas at the BS and two receive antennas at each UE The main simulation parameters and assumptions are summarized in Table 1.

The RRM functionalities are controlled by the packet scheduler together with link adaptation and HARQ entities. Link adaptation consists of two separate elements – the inner loop LA (ILLA) and the outer loop LA (OLLA). These are used for removing CQI imperfections, estimating supported data rates and MCS's, and stabilizing the 1st transmission Block Error Probability (BLEP) to the target range (typically 10-20%). Simple admission control scheme is used for keeping the number of UEs per cell constant. HARQ is based on SAW protocol and a maximum number of three re-transmissions is allowed. MIMO functionality requires individual HARQ entry per stream which is also implemented. Link-to-system level mapping is based on the effective SINR mapping (EESM) principle [1].

As a concrete example of unequal PRB power allocation we exploit the case of reducing the transmit power of every third RB by 1dB and the neighbouring ones yet by another 1dB (i.e. the relative power pattern is 0dB, -1dB, -2dB, 0dB, -1dB, -2dB, …) compared to maximum transmit power [18]. Notice that this is just one concrete example selected for evaluating and comparing the performance of different schedulers.

SFR scheme benefits from full bandwidth utilization for each BS and sub-band division helps in mitigating the CCI. The reuse factor used in the simulation scenarios is 3, which corresponds to sub-band division of (17, 17, 16) RB's (total of 50 RB's). Two different power patterns i.e. SFR power masks (described in section 2) define the power levels used for each RB group. In the first case the corresponding power pattern is (0dB, -4dB, -4dB) and in second case (0dB, -1dB, -4dB).

Table 1: Default simulation parameters

| Parameter | Assumption |
|---|---|
| Cellular Layout | Hexagonal grid, 19 cell sites, 3 sectors per site |
| Inter-site distance | 500 m |
| Carrier Frequency / Bandwidth | 2000MHz / 10 MHz |
| Channel estimation | Ideal |
| PDP | ITU Typical Urban 20 paths |
| Minimum distance between UE and cell | >= 35 meters |
| Average number of UEs per cell | 15 |
| Max. number of frequency multiplexed UEs | 10 |
| UE receiver | MRC and LMMSE |
| Shadowing standard deviation | 8 dB |
| UE speed | 3km/h |
| Total BS TX power ($P_{total}$) | 46dBm - 10MHz carrier |
| Traffic model | Full Buffer |
| Fast Fading Model | Jakes Spectrum |
| CQI reporting time | 5 TTI |
| CQI delay | 2 TTIs |
| MCS rates | QPSK (1/3, 1/2, 2/3),16QAM(1/2, 2/3, 4/5),64QAM(1/2, 2/3, 4/5) |
| ACK/NACK delay | 2ms |
| Number of SAW channels | 6 |
| Maximum number of retransmissions | 3 |
| HARQ model | Ideal CC |
| 1st transmission BLER target | 20% |
| Forgetting factor | 0.002 |
| Scheduling schemes | PF, PPF,MPMPF |

The actual effective SINR calculations rely on subcarrier-wise complex channel gains (estimated using reference symbols in practice) and depend in general also on the assumed receiver (detector) topology. Here we assume that the LMMSE detector, properly tailored for the transmission mode (1-stream SU, 2-stream SU or 2-stream MU) is deployed. The detector structures and SINR modelling for different transmission modes are described in detail below.

### 4.2. Detectors and SINR Modeling

#### 4.2.1. Single-Stream SU Case

In this case, only one of the two BS transmit antennas is used to transmit one stream. At individual time instant (time-index dropped here), the received spatial 2x1 signal vector of UE $i$ at sub-carrier $c$ is of the form

$$\mathbf{y}_{i,c} = \mathbf{h}_{i,c} x_{i,c} + \mathbf{n}_{i,c} + \mathbf{z}_{i,c} \tag{3}$$

where $x_{i,c}$, $\mathbf{h}_{i,c}$, $\mathbf{n}_{i,c}$ and $\mathbf{z}_{i,c}$ denote the transmit symbol, 2x1 channel vector, 2x1 received noise vector and 2x1 inter-cell interference vector, respectively. Then the LMMSE detector $\hat{x}_{i,c} = \mathbf{w}_{i,c}^H \mathbf{y}_{i,c}$ is given by

$$\mathbf{w}_{i,c} = \sigma_{x,i}^2 \sigma_{i,c}^2 (\mathbf{h}_{i,c}^H \sigma_{x,i}^2 \sigma_{i,c}^2 \mathbf{h}_{i,c} + \mathbf{\Sigma}_{n,i} + \mathbf{\Sigma}_{z,i})^{-1} \mathbf{h}_{i,c} \tag{4}$$

where $\sigma_{x,i}^2, \sigma_{i,c}^2$, $\boldsymbol{\Sigma}_{n,i}$ and $\boldsymbol{\Sigma}_{z,i}$ denote the transmit power (per the used antenna), power mask per sub-carrier, noise covariance matrix and inter-cell interference covariance matrix, respectively. Now the SINR is given by

$$\gamma_{i,c} = \frac{\left|\mathbf{w}_{i,c}^H \mathbf{h}_{i,c}\right|^2 \sigma_{x,i}^2 \sigma_{i,c}^2}{\mathbf{w}_{i,c}^H \boldsymbol{\Sigma}_{n,i} \mathbf{w}_{i,c} + \mathbf{w}_{i,c}^H \boldsymbol{\Sigma}_{z,i} \mathbf{w}_{i,c}} \quad (5)$$

The noise variables at different receiver antennas are assumed uncorrelated (diagonal $\mathbf{S}_{n,i}$) while the more detailed modeling of inter-cell interference (structure of $\mathbf{S}_{z,i}$) takes into account the distances and channels from neighboring base stations (for more details, see e.g. [17]).
The MRC detector is given by

$$\mathbf{w}_{i,c} = \frac{\mathbf{h}_{i,c}}{\|\mathbf{h}_{i,c}\|^2} \quad (6)$$

### 4.2.2. Dual-Stream SU Case

In this case, both of the two BS transmit antennas are used for transmission, on one stream per antenna basis. At individual time instant, the received spatial 2x1 signal vector of UE $i$ at sub-carrier $c$ is now given by

$$\mathbf{y}_{i,c} = \mathbf{H}_{i,c} \mathbf{x}_{i,c} + \mathbf{n}_{i,c} + \mathbf{z}_{i,c} \quad (7)$$

where $\mathbf{x}_{i,c}$ and $\mathbf{H}_{i,c} = [\mathbf{h}_{i,c,1}, \mathbf{h}_{i,c,2}]$ denote the 2x1 transmit symbol vector and 2x2 channel matrix, respectively. Now the LMMSE detector $\hat{\mathbf{x}}_{i,c} = \mathbf{W}_{i,c} \mathbf{y}_{i,c}$ is given by

$$\mathbf{W}_{i,c} = \boldsymbol{\Sigma}_{x,i} \mathbf{H}_{i,c}^H (\mathbf{H}_{i,c} \boldsymbol{\Sigma}_{x,i} \mathbf{H}_{i,c}^H + \boldsymbol{\Sigma}_{n,i} + \boldsymbol{\Sigma}_{z,i})^{-1} = \begin{bmatrix} \mathbf{w}_{i,c,1}^H \\ \mathbf{w}_{i,c,2}^H \end{bmatrix} \quad (8)$$

where $\boldsymbol{\Sigma}_{x,i} = \sigma_{i,c}^2 diag\{\sigma_{x,i,1}^2, \sigma_{x,i,2}^2\} = \sigma_{i,c}^2 diag\{\sigma_{x,i}^2/2, \sigma_{x,i}^2/2\}$ denotes the 2x2 covariance matrix (assumed diagonal) of the transmit symbols. Note that compared to single-stream case, the overall BS transmit power is now divided between the two antennas, as indicated above. Then the SINR's for the two transmit symbols are given by

$$\gamma_{i,c,1} = \frac{\left|\mathbf{w}_{i,c,1}^H \mathbf{h}_{i,c,1}\right|^2 \sigma_{x,i,1}^2 \sigma_{i,c}^2}{\left|\mathbf{w}_{i,c,1}^H \mathbf{h}_{i,c,2}\right|^2 \sigma_{x,i,2}^2 \sigma_{i,c}^2 + \mathbf{w}_{i,c,1}^H \boldsymbol{\Sigma}_{n,i} \mathbf{w}_{i,c,1} + \mathbf{w}_{i,c,1}^H \boldsymbol{\Sigma}_{z,i} \mathbf{w}_{i,c,1}}$$

$$\gamma_{i,c,2} = \frac{\left|\mathbf{w}_{i,c,2}^H \mathbf{h}_{i,c,2}\right|^2 \sigma_{x,i,2}^2 \sigma_{i,c}^2}{\left|\mathbf{w}_{i,c,2}^H \mathbf{h}_{i,c,1}\right|^2 \sigma_{x,i,1}^2 \sigma_{i,c}^2 + \mathbf{w}_{i,c,2}^H \boldsymbol{\Sigma}_{n,i} \mathbf{w}_{i,c,2} + \mathbf{w}_{i,c,2}^H \boldsymbol{\Sigma}_{z,i} \mathbf{w}_{i,c,2}} \quad (9)$$

### 4.2.3. Dual-Stream MU Case

In this case, the transmission principle and SINR modelling are similar to subsection 2) above, but the two spatially multiplexed streams belong now to two different UE's, say $i$ and $i'$. Thus the SINR's in (9) are interpreted accordingly.

Finally, for link-to-system level mapping purposes, the exponential effective SINR mapping (EESM), as described in [1-3], is deployed.

## 5. NUMERICAL RESULTS AND ANALYSIS

This section presents the obtained results from the radio system simulations using different PS algorithms combined with SFR schemes as described in the paper. The system-level performance is generally measured and evaluated in terms of:
- Throughput - the total number of successfully delivered bits per unit time. Usually measured either in kbps or Mbps.
- Coverage – the experienced data rate per UE at the 95% coverage probability.
- Fairness per scheduling scheme measured using Jain's fairness index [19]

Initially, we demonstrate the behaviour of the proposed MPMPF scheduler by using different power coefficients $α_1$ and $α_2$ and comparing it against other PF scheduling algorithms. Secondly, we illustrate the potential of combining the use of different SFR power masks and again tuning the MPMPF scheduler' power coefficients. To emphasize the role of power-aware and CQI based priority metric calculation in (2), we fix the value of $α_2$ to 1 and change the values of $α_1$ as $α_1 = \{1,2,4\}$ [11], [12].

Figure 3 illustrates the average cell throughput and cell-edge coverage for the different schedulers in SIMO (sub-figures (a) and (b)) and MIMO (sub-figures (c) and (d)) system simulation cases. The power coefficient values are presented as index M, where M1 represents the first couple, i.e., $α_1=1$, $α_2=1$, M2: $α_1=2$, $α_2=1$ and M3: $α_1=4$, $α_2=1$. The obtained results with the proposed scheduler are compared with the reference PF schedulers – ordinary PF, power-aware PF described in (1) and MMPF scheduler from [13]. For the cases M1-M3, based on Figure 2, the new MPMPF scheduler achieves coverage gains in the order of 63-91% at the expense of only 15-22% throughput loss compared to ordinary PF in the SIMO scenario (sub-figures (a) and (b)). In the corresponding MIMO (sub-figures (c) and (d)) scenario, we obtain 71% constant coverage gain for the same throughput loss as in previous SIMO case. Compared to the MMPF scheduling principle [16] or to the power-aware PF in (1), similar coverage gains are obtained, as can be read from the figure.

Similarly, Figure 4 illustrates the same performance statistics as in Figure 3 for the different scheduling approaches and different power mask cases. By combining SFR and PF scheduling we achieve small throughput gain, while coverage is increased with 10%. Changing the scheduling method to MPMPF and coefficient values (M1), we achieve coverage gains in the order of 58% at the expense of 20% throughput loss for PM1 case scenario presented in (a) and (b). In PM2 case scenario illustrated on (c) and (d) we obtain nearly the same coverage gain (60%) for small throughput loss (13%).

Continuing the evaluation of the proposed method – SFR combined with MPMPF, we clearly see a trade-off between average cell throughput and coverage for different power coefficient values in PM1 and PM2 cases. Furthermore, the remaining power coefficient values can be used for tuning the overall scheduling performance. For the rest of the cases, the cell throughput loss is decreased stepwise with around 4% per index M, which on the other hand corresponds to coverage gains increase from 74% to 80%. Consequently, an obvious trade-off between average cell throughput and coverage is clearly seen. Similarly, in PM2 case scenario, we obtain coverage gains between 75% to 82% corresponding to throughput losses from 10% to 8% as illustrated by the performance statistics.

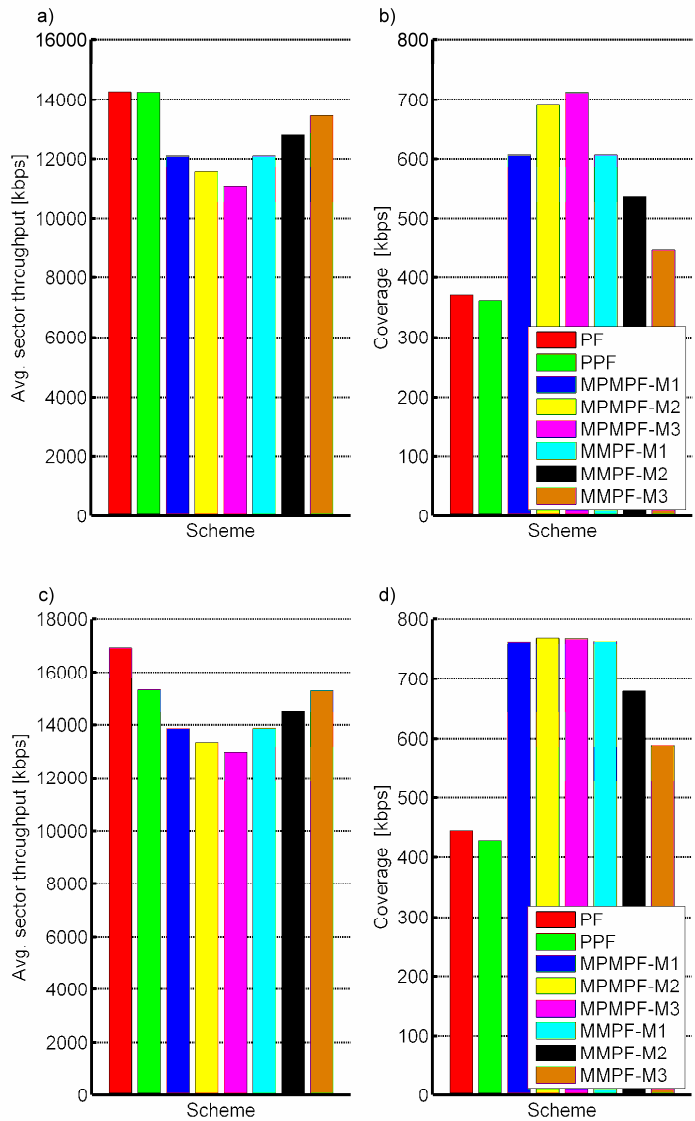

Figure 3: Average cell throughput and coverage gain over the reference PF scheduling scheme for the different simulation cases – SIMO (a, b) and MIMO (c, d).

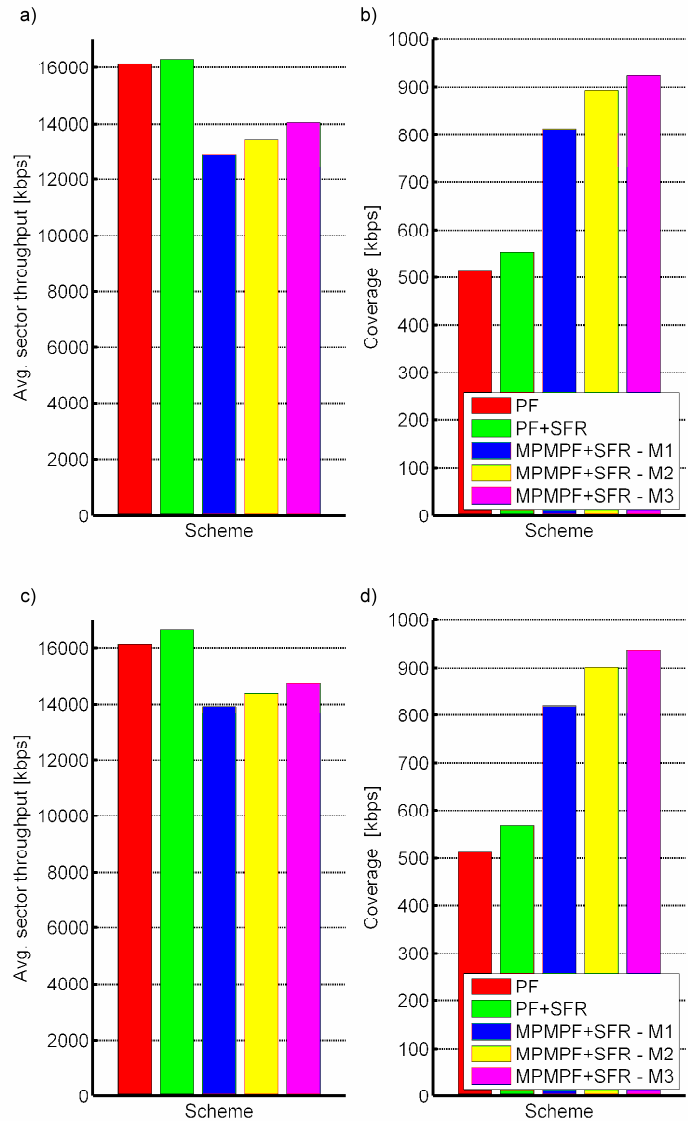

Figure 4: Average cell throughput and coverage gain over the reference PF scheduling scheme and PF +SFR scheme for the different power masks scenarios – PM1 (a, b) and PM2 (c, d). The schemes M1-M3 refer to the MPMPF scheduler with power coefficient values.

Figure 5 illustrates the Jain's fairness index [14] per scheduling scheme for SIMO and MIMO scenarios, calculated using the truly realized throughputs at each TTI for all 15 UE's and over all the simulation runs. The value on the x axis corresponds to the used scheduler type (1 refers to ordinary PF scheduler, 2 refers to PPF, etc.). Clearly, the proposed MPMPF scheduler outperforms all other scheduling algorithms and the received fairness gains are in range of 17%-20% in the SIMO case and 35%-37% in the MIMO case.

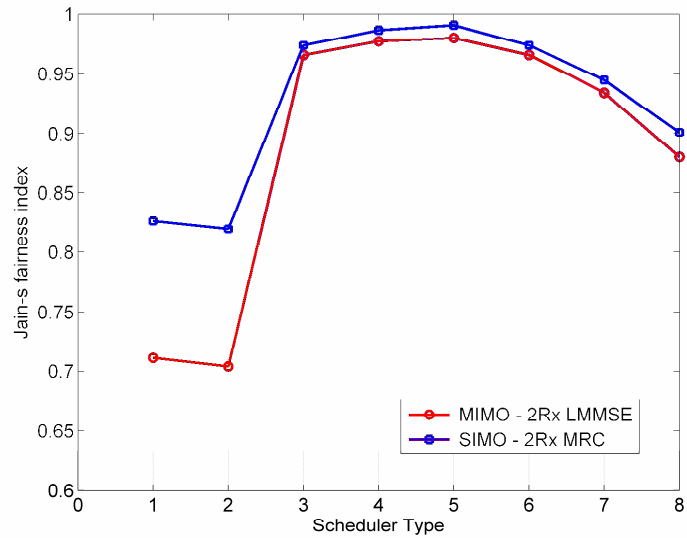

Figure 5: Jain's fairness index per scheduling scheme for SIMO and MIMO cases. Scheduler type 1 means ordinary PF, 2 means power-aware PF from (1), 3-5 mean proposed modified power-aware PF from (2) and 6-8 mean modified PF from [13].

Figure 6 illustrates the Jain's fairness index per scheduling scheme for different SFR power masks scenarios, calculated over all the $I_{TOT} = 15$ UE's. The value on the x axis corresponds to used scheduler type, where 1 refers to the reference PF scheduler, 2 refers to PF+SFR, 3 refers to the MPMPF+SFR with index M1, etc. Clearly, the fairness distribution with MPMPF+SFR outperforms the used reference plain PF scheduler and PF+SFR for both analyzed cases. The received fairness gains are in range of 17% to 31% in the PM1 case and 18% to 32% in the PM2 case. Compared to the PF+SFR case the corresponding gains are in the range of 15% to 17% for both cases.

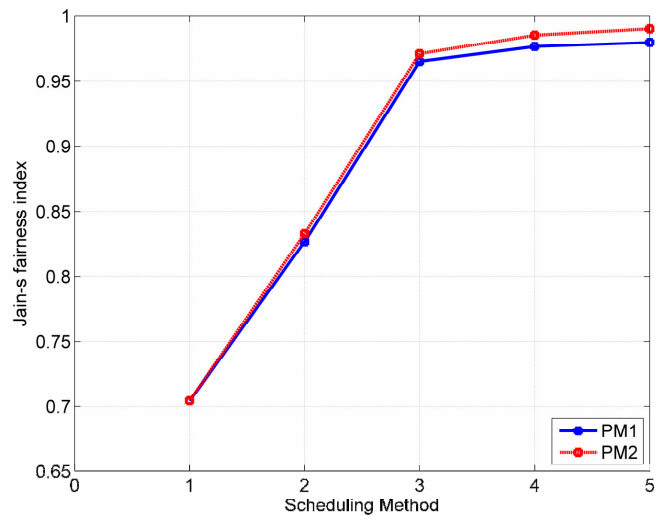

Figure 6: Jain's fairness index per scheduling scheme. Scheduler type 1 means ordinary PF, while 2 means PF +SFR, 3-5 means MPMPF +SFR with power coefficients M1,M2 and M3 correspondingly.

## 6. CONCLUSIONS

In this paper, a new power-aware multi-stream proportional fair scheduling metric covering time-, frequency- and spatial domains was proposed. Moreover, we have studied the potential of combining soft frequency reuse schemes with advanced power-aware multi-user packet scheduling algorithms in OFDMA type radio system context, using UTRAN long term evolution (LTE) downlink in Macro cell environment as a practical example. The proposed MPMPF metric takes into account the transmit power allocation on RB basis defined by SFR power masks, the instantaneous channel qualities (CQI's) as well as resource allocation fairness. The achievable throughput performance together with coverage and fairness distributions were analyzed and compared against the corresponding statistics of more traditional earlier-reported proportional fair scheduling techniques with SDM functionality in SIMO and MIMO cases, as well as combined with SFR functionality. In the case of fixed coverage requirements, the proposed scheduling metric calculations based on UE channel feedback offers better control over the ratio between the achievable cell/UE throughput and coverage increase, as well as increased UE fairness taking into account irregular BS transmission power. As a practical example, the fairness in resource allocation together with cell coverage can be increased significantly (more than 50%) by allowing a small decrease (in the order of only 10-15%) in the cell throughput for plain PF scheduling case and more than 15% increase over the traditional PF when power consideration are taken into account or combined with SFR.

## ACKNOWLEDGEMENTS

The authors would like to thank Prof. Jukka Lempiäinen, Tampere University of Technology, Tampere, Finland, for fruitful discussions.

**Authors**

**Stanislav Nonchev** was born in Burgas, Bulgaria, on January 14, 1981. He received the M.Sc. Degree in telecommunications engineering (TE) from Tampere University of Technology (TUT), Finland, in 2006. Currently he is a Ph.D student at Tampere University of Technology, Department of Communications Engineering. His research spans radio resource management for ad-hoc and mobile networks, radio network planning, and network protocols.

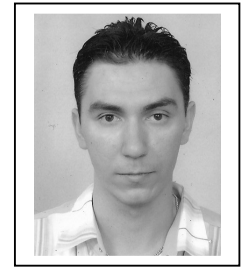

**Mikko Valkama** was born in Pirkkala, Finland, on November 27, 1975. He received the M.Sc. and Ph.D. Degrees (both with honors) in electrical engineering (EE) from Tampere University of Technology (TUT), Finland, in 2000 and 2001, respectively. In 2002 he received the Best Ph.D. Thesis -award by the Finnish Academy of Science and Letters for his thesis entitled "Advanced I/Q signal processing for wideband receivers: Models and algorithms". In 2003, he was working as a visiting researcher with the Communications Systems and Signal Processing Institute at SDSU, San Diego, CA. Currently, he is a Full Professor at the Department of Communications Engineering at TUT, Finland. He has been involved in organizing conferences, like the IEEE SPAWC'07 (Publications Chair) held in Helsinki, Finland. His general research interests include communications signal processing, estimation and detection techniques, signal processing algorithms for software defined flexible radios, digital transmission techniques such as different variants of multicarrier modulation methods and OFDM, and radio resource management for ad-hoc and mobile networks.

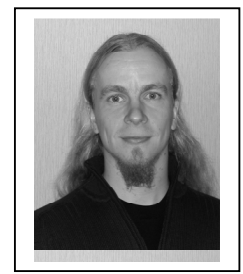